\documentclass[a4paper]{easychair}

\usepackage{amssymb}
\usepackage{amsmath}
\usepackage{amsthm}
\usepackage{etex}
\usepackage{ifthen}
\usepackage{bussproofs}
\usepackage{caption}
\usepackage[noadjust]{cite}
\usepackage{graphicx}
\usepackage{mdwlist}
\usepackage{subfig}
\usepackage{floatrow}
\usepackage{ulem}
\usepackage{tikz}
\usetikzlibrary{positioning,fit,arrows,decorations.markings}
\usepackage{xspace}
\usepackage{fancybox}
\usepackage{booktabs}
\usepackage{etex}
\definecolor{linkcolor}{rgb}{0.2,0.2,0.7}%
\hypersetup{%
  pdftex,%
  colorlinks=true,%
  pdfborder={0 0 0},%
  linkcolor=linkcolor,%
  citecolor=linkcolor,%
  urlcolor=linkcolor,%
}

\newcommand{\IFF}{iff\xspace}
\newcommand{\auxeqs}{auxiliary equations\xspace}
\newcommand{\cstep}{\leadsto}
\newcommand\COMPrulename[2][]{%
  \ensuremath{\textsf{#2}\ifthenelse{\equal{#1}{}}{}{_\textsf{#1}}}\xspace}
\newcommand\orientl{\COMPrulename[l]{orient}}
\newcommand\orientr{\COMPrulename[r]{orient}}
\newcommand\simplifyl{\COMPrulename[l]{simplify}}
\newcommand\simplifyr{\COMPrulename[r]{simplify}}
\newcommand\delete{\COMPrulename{delete}}
\newcommand\compose{\COMPrulename{compose}}
\newcommand\collapse{\COMPrulename{collapse}}
\newcommand\deduce{\COMPrulename{deduce}}
\newcommand\irule[3][]{%
  \frac{\strut\displaystyle#2}{\strut\displaystyle#3}%
  \ifthenelse{\equal{#1}{}}{}{\,{\scriptstyle(#1)}}}
\newcommand\fs[1]{\mathsf{#1}}
\newcommand{\convert}{\leftrightarrow}
\newcommand{\from}{\leftarrow}
\newcommand\ifrom[1]{\smash{\stackrel{#1}{\from}}}
\newcommand\ito[1]{\smash{\stackrel{#1}{\to}}}
\def\test#1#2#3{\setbox0=\hbox{$\vphantom{#1}^{#2}_{#3}$}%
                \dimen0=\wd0%
                \setbox1=\hbox{$\scriptstyle #2$}%
                \advance\dimen0-\wd1%
                \setbox1=\hbox{\hskip\dimen0\copy1}%
                \dimen0=\wd0%
                \setbox2=\hbox{$\scriptstyle #3$}%
                \advance\dimen0-\wd2%
                \setbox2=\hbox{\hskip\dimen0\copy2}%
                {\vphantom{#1}^{\box1}_{\box2}}{#1}
}
\newcommand{\TT}{{\cal T}}
\newcommand{\FromTBA}[3]{\mathrel{\test{#3}{#1}{#2}}}
\newcommand{\FromTB}[2]{\FromTBA{#1}{#2}{\from}}
\newcommand\SIG[1]{\ensuremath{\mathcal{#1}}}
\newcommand\VAR[1]{\ensuremath{\mathcal{#1}}}
\newcommand\FF{\SIG{F}}
\newcommand\VV{\VAR{V}}
\newcommand\EE{\ensuremath{\mathcal{E}}\xspace}
\newcommand\HH{\ensuremath{\mathcal{H}}\xspace}
\newcommand\RR{\ensuremath{\mathcal{R}}\xspace}
\newcommand{\rTH}[1]{Theorem~\ref{#1}}
\newcommand{\rTB}[1]{Table~\ref{#1}}
\newcommand\rST[2]{Table~\mbox{\ref{#1}\subref{#2}}}
\newcommand{\rEX}[1]{Example~\ref{#1}}
\newcommand\rFG[1]{Figure~\ref{#1}}
\newcommand\CIME{\textsf{C}\textit{i}\textsf{ME}3\xspace}
\newcommand\KBCV{\textsf{KBCV}\xspace}
\newcommand\MKBTT{\textsf{MKB{\scalebox{0.9}{TT}}}\xspace}
\newcommand\isafor{\textsf{Isa\kern-0.2exF\kern-0.2exo\kern-0.2exR}\xspace}
\newcommand\ceta{\textsf{C\kern-0.2exe\kern-0.5exT\kern-0.5exA}\xspace}
\newcommand\phase[1]{\emph{#1}}
\newcommand\file[1]{\nolinkurl{#1}}

\title{Recording Completion for
Finding and Certifying Proofs in Equational Logic%
\footnote{%
  Supported by Austrian Science Fund (FWF): P22467, P22767, J3202,
  and a grant by Hypo Tirol Bank.}}
\titlerunning{Recording Completion}

\author{%
  Thomas Sternagel, Ren{\'e} Thiemann, Harald Zankl\\[-0.5ex]
  \affiliation{Institute of Computer Science}\\
  \affiliation{University of Innsbruck, Austria}\\[1ex]
\and
  Christian Sternagel\\[-0.5ex]
  \affiliation{School of Information Science}\\
  \affiliation{JAIST, Japan}\\
}
\authorrunning{T.~Sternagel et al.}

\volumeinfo
  {???}
  {1}
  {submitted to 1st International Workshop on Confluence}
  {1}
  {1}
  {1}

\theoremstyle{plain}
\newtheorem{theorem}{Theorem}
\newtheorem{lemma}{Lemma}
\newtheorem{corollary}{Corollary}

\theoremstyle{definition}
\newtheorem{example}{Example}
\newtheorem{definition}{Definition}

\begin{document}
\maketitle
 
\section{Introduction}

Solving the word problem 
requires
to decide whether an equation $s \approx t$ follows from an equational
system~(ES)~\EE. By Birkhoff's theorem this is equivalent to the existence of 
a conversion $s \convert^*_\EE t$.  Knuth-Bendix
completion~\cite{KB70} (if successful) gives a decision procedure:
If an ES~\EE is transformed into an equivalent convergent
term rewrite system~(TRS)~\RR, then $s \convert^*_\EE t$ \IFF the \RR-normal
forms of $s$ and $t$ coincide.
(Note that completion does not construct such conversions explicitly.)

\begin{example} 
\label{EX:1}
For~$\EE = \{
\fs{ff} \approx\fs{f},
\fs{ggf} \approx\fs{g}
\}$
(where $\fs{f}$ and $\fs{g}$ are unary function symbols, for which we find it
convenient to abbreviate $\fs{f}(\fs{g}(\fs{f}(x)))$ to $\fs{fgf}$ etc.) 
a possible choice of \RR is 
$\{
\fs{ff} \to \fs{f},
\fs{gf} \to \fs{g},
\fs{gg} \to \fs{g}
\}$.
Since
$\fs{fgf} \to^*_\RR \fs{fg} \FromTB{*}{\RR} \fs{fgg}$, we have that
$\fs{fgf} \approx \fs{fgg}$ follows from~\EE.
\end{example}

When we want to answer/certify whether $s \convert^*_\EE t$,
we 
face the following situation:
(1)
It is hard to find a conversion but
easy to certify a given one.
(2)
Under the assumption that Knuth-Bendix completion is successful, it is easy to
decide the existence of a conversion (just rewrite $s$ and $t$
to \RR-normal forms) but hard to certify this decision (e.g., by certifying that
\EE and \RR are equivalent).

In this paper we introduce \emph{recording completion}, which overcomes
both problems. Recording completion keeps a history that allows us to reconstruct
how the rules in~\RR have been derived from the equations in~\EE. Then, from a join
$s \to^*_\RR \cdot \FromTB{*}{\RR} t$ a conversion
$s \convert^*_\EE t$ can be reconstructed. 
Furthermore, recording completion
enables the certification of completion proofs,
i.e., to check that~\RR and~\EE are equivalent.
Using equivalence together with confluence and termination certificates, 
it is also possible to certify 
that a conversion $s \convert^*_\EE t$ does \emph{not} exist.

In addition to formalizing
all required theorems like the critical pair theorem and
soundness of completion,
we have proven two new results: For 
finite completion proofs, i.e., where the completion procedure 
stops successfully after a finite number of steps, the 
strict encompassment condition
(in the \collapse-rule of \rFG{FIG:rcir}) is not required.
Moreover, an infinite set of variables
is essential for the critical pair theorem as well as modularity of confluence
\cite{Toyama87}.

\section{Proof Construction via Recording Completion}
\label{SEC:rec}

We extend the inference rules of completion
\cite{BN98} by a \emph{history component} which allows us to infer
how rules in~\RR have been derived from equations in~\EE.
The construction of a conversion ${s \convert^*_\EE t}$ (if
possible at all) is then executed in three phases:
(\textbf{record})
The inference rules of recording completion (see \rFG{FIG:rcir})
are applied to the ES~\EE. Upon success, a convergent
TRS~\RR (equivalent to~\EE) and a history~\HH (recording how
the rules in~\RR have been derived) are computed. 
(\textbf{compare})
If the previous phase is successful, 
the test for ${s \to^*_\RR \cdot \FromTB{*}{\RR} t}$ is performed.
(\textbf{recall})
If the previous phase is successful, we construct
$s \convert^*_\EE t$ from ${s \to^*_\RR \cdot \FromTB{*}{\RR}t}$.

\pagebreak
In the sequel we give more details for each of the phases.

\paragraph{Record.}
\begin{figure}[tb]
  \begin{center}
  $\begin{array}{c@{\qquad}l}
  \irule[\deduce]%
    { (\EE,\RR,\HH)}%
    {(\EE\cup\{m\colon  s\approx t\},\RR,%
      \HH\cup\{m\colon s\stackrel{j}{\leftarrow}u\stackrel{k}{\to}t\})}
    &\text{if $s\FromTBA{}{\RR}{\stackrel{j}{\leftarrow}} u\stackrel{k}{\to}_\RR t$}
  \\[3ex]
  \irule[\orientl]%
    {(\EE\cup\{i\colon  s\approx t\},\RR,\HH)}%
    {(\EE,\RR\cup\{i\colon  s\to t\},\HH)}%
  &\text{if $s > t$}
  \\[2ex]
  \irule[\orientr]%
    {(\EE\cup\{i\colon  s\approx t\},\RR,%
      \HH\cup\{i\colon s\stackrel{j}{\circ_1}u\stackrel{k}{\circ_2}t\})}%
    {(\EE,\RR\cup\{i\colon  t\to s\},%
      \HH\cup\{i\colon t\mathop{(\stackrel{k}{\circ_2})^{-1}}u\mathop{(\stackrel{j}{\circ_1})^{-1}}s\})}
    &\text{if $t > s$}
  \\[3ex]
  \irule[\simplifyl]%
    {(\EE\cup\{i\colon s \approx t\},\RR,\HH)}%
    {(\EE\cup\{m\colon  u \approx t\},\RR,%
      \HH\cup\{m\colon  u\stackrel{l}{\leftarrow}s\stackrel{i}{\to}t\})}
    &\text{if $s\stackrel{l}{\to}_\RR u \hfill$}
  \\[3ex]
  \irule[\simplifyr]%
    {(\EE\cup\{i\colon  s\approx t\},\RR,\HH)}%
    {(\EE\cup\{m\colon  s \approx u\},\RR,%
      \HH\cup\{m\colon  s\stackrel{i}{\to}t\stackrel{l}{\to}u\})}
    &\text{if $t\stackrel{l}{\to}_\RR u\hfill$}
  \\[3ex]
  \irule[\delete]%
    {(\EE\cup\{i\colon  s\approx s\},\RR,%
      \HH\cup\{i\colon s\circ_1 v \circ_2 s\})}%
    {(\EE,\RR,\HH)}
  \\[3ex]
  \irule[\compose]%
    {(\EE,\RR\cup\{i\colon  s\to t\},\HH)}%
    {(\EE,\RR\cup\{m\colon s\to u\},%
      \HH\cup\{m\colon s\stackrel{i}{\to}t\stackrel{j}{\to}u\})}
    &\text{if $t\stackrel{j}{\to}_\RR u$}
  \\[4ex] 
  \irule[\collapse]%
    {(\EE,\RR\cup\{i\colon  s\to t\},\HH)}%
    {(\EE\cup\{m\colon u\approx t\},\RR,%
      \HH\cup\{m\colon u\stackrel{j}{\leftarrow}s\stackrel{i}{\to}t\})}
 & \text{if $s\stackrel{j}{\to}_\RR u$}
  \end{array}$
  \end{center}
  \mbox{}\\
  \caption{The inference rules of \emph{recording completion}.}
  \label{FIG:rcir}
\vspace{-2ex}
\end{figure}

The \phase{record} phase uses the inference rules from \rFG{FIG:rcir}
where every rule/equation is annotated by a unique index $i$. Here,
$\smash{\stackrel{i}\to_\RR}$ denotes an $\RR$-reduction using the rule with
index $i$.
The inference rules are similar to the standard rules except 
for the following two differences:
In the $\collapse$-rule we dropped the condition of strict
encompassment. Since we only consider finite runs,
this condition is no longer required for soundness (cf.\ \rTH{completion thm}).
Furthermore, there is a new history component $\HH$ whose entries are of the form
$\smash{i: s\stackrel{j}{\circ_1}u\stackrel{k}{\circ_2}t}$ where $i$ is the
index of the
entry, $j$ and $k$ are indices of equations/rules, $s$, $u$, and $t$
are terms, and $\circ_1, \circ_2 \in \{{\leftarrow}, {\to}, {\approx}\}$.

Let us take a closer look at the extended inference rules.
For \deduce the peak $\smash{s \stackrel{j}{\leftarrow}_\RR u \stackrel{k}{\to}_\RR t}$
that triggers the new equation $s \approx t$
is stored in a history entry (where $m$ is assumed to be a fresh index
that is larger than every earlier index).
By \orientl we orient an equation from left to right and the
corresponding history entry remains unchanged, whereas
by \orientr we orient an equation from right to left and
thus have to ``mirror'' the corresponding history entry.
Here $>$ is a reduction order, which is part of the input.
The rules \simplifyl and \simplifyr are used to \RR-rewrite
a left- or right-hand side of an equation.
With \delete we remove trivial
equations from~\EE and the corresponding history entry from~\HH.
Finally, \compose rewrites a right-hand side of a rule in \RR while
\collapse does the same for left-hand sides.

We write
$(\EE_i,\RR_i,\HH_i) \cstep (\EE_{i+1},\RR_{i+1},\HH_{i+1})$
for the application of an arbitrary inference rule to the triple
$(\EE_i,\RR_i,\HH_i)$
resulting in
$(\EE_{i+1},\RR_{i+1},\HH_{i+1})$.

\begin{definition}
A \emph{run} of recording completion for $\EE$ is a finite sequence
$(\EE_0, \RR_0, \HH_0) \cstep^n (\EE_n, \RR_n, \HH_n)$ of rule applications,
where
$\EE_0 = \{i : s \approx t \mid s \approx t \in \EE\}$ with fresh index $i$ for
each equation,
$\RR_0 = \varnothing$, and the initial history is 
$\HH_0 = \{i: \smash{s\stackrel{i}{\to}t\stackrel{0}{\approx} t \mid i : s\approx t \in
\EE_0}\}$.
A run is \emph{successful} if
$\EE_n = \varnothing$ and all critical pairs of $\RR_n$ that are not contained
in $\bigcup_{i \leq n} \EE_i$ are joinable by $\RR_n$.
A run is \emph{sound} if~$\RR_n$ is convergent and equivalent to~$\EE_0$.
\end{definition}

The requirement on critical pairs for a successful run can be replaced by local confluence
of $\RR_n$.

\begin{example}\label{EX:rc}
\mbox{}
\begin{table}%
\floatbox{table}[\textwidth]{%
  \caption{\label{TAB:rc}Example of recording completion.}%
}{\begin{subfloatrow}
\subfloat[\label{TAB:rcis} Initial state.]{
\begin{tabular}{rlcrl}
\toprule
\multicolumn{2}{c}{$\EE_0$}
  & $\RR_0$
  & \multicolumn{2}{c}{$\HH_0$}
\\ \midrule
1:&$\fs{ff}\approx\fs{f}$
 & $\varnothing$
 & 1:&$\fs{ff}\stackrel{1}{\to}\fs{f} \approx \fs{f}$%
\\
2:&$\fs{ggf}\approx\fs{g}$
 &
 & 2:&$\fs{ggf}\stackrel{2}{\to}\fs{g} \approx \fs{g}$%
\\
\bottomrule
\end{tabular}
}
\qquad
\subfloat[\label{TAB:rcfs} Final state.]{%
\begin{tabular}{crlrl}
\toprule
$\EE_n$
  & \multicolumn{2}{c}{$\RR_n$}
  & \multicolumn{2}{c}{$\HH_n$}
\\ \midrule
$\varnothing$&%
1:&$\fs{ff}\to\fs{f}$&%
1:&$\fs{ff}\stackrel{1}{\to}\fs{f} \stackrel{0}{\approx} \fs{f}$%
\\
&%
4:&$\fs{gf}\to\fs{g}$&%
2:&$\fs{ggf}\stackrel{2}{\to}\fs{g} \stackrel{0}{\approx} \fs{g}$%
\\
&%
5:&$\fs{gg}\to\fs{g}$&%
3:&$\fs{ggf}\stackrel{1}{\leftarrow}\fs{ggff}\stackrel{2}{\to}\fs{gf}$%
\\
&%
&&%
4:&$\fs{gf}\stackrel{3}{\leftarrow}\fs{ggf}\stackrel{2}{\to}\fs{g}$%
\\
&%
&&%
5:&$\fs{gg}\stackrel{4}{\leftarrow}\fs{ggf}\stackrel{2}{\to}\fs{g}$%
\\
\bottomrule
\end{tabular}
}%
\end{subfloatrow}}%
\end{table}
Recall $\EE$ from \rEX{EX:1}.
We start with the triple depicted in \rST{TAB:rc}{TAB:rcis} and perform recording
completion.
Note that LPO with empty precedence orients all emerging rules in the
desired direction.
After orienting rules~$1$ and~$2$ from left to right we deduce a critical pair
between rules~2 and~1, resulting in the equation
$3: \fs{ggf} \approx \fs{gf}$ and the history entry
$3: \fs{ggf} \ifrom{1} \fs{ggff} \ito{2} \fs{gf}$.
Next we simplify the left-hand side of equation~$3$ by an application of
rule~$2$ and obtain the equation $4: \fs{g} \approx \fs{gf}$ with
corresponding history entry
$4: \fs{g} \ifrom{2} \fs{ggf} \ito{3} \fs{gf}$.
Orienting rule $4$ from right to left causes the history entry to be
mirrored, i.e., 
$4: \fs{gf} \ifrom{3} \fs{ggf} \ito{2} \fs{g}$.
Rules~$2$ and~$4$ allow to deduce equation
$5: \fs{gg} \approx \fs{g}$ with history
$5: \fs{gg} \ifrom{4} \fs{ggf} \ito{2} \fs{g}$, which we orient from left
to right. Collapsing the left-hand side of rule~$2$ with rule~$5$ yields
$6: \fs{gf} \approx \fs{g}$ with $6: \fs{gf} \ifrom{5} \fs{ggf} \ito{2} \fs{g}$.
Now rule~$4$ simplifies equation~$6$ into
$7: \fs{g} \approx \fs{g}$ with $7: \fs{g} \ifrom{4} \fs{gf} \ito{6} \fs{g}$,
which is immediately deleted afterwards.
Finally, $\EE_n$ is empty and 
as all remaining critical pairs of $\RR_n$ are joinable,
the procedure can be stopped.
Since there is no rule with index~$6$, the history entry~$6$ can be dropped.
Hence, we obtain the result depicted in \rST{TAB:rc}{TAB:rcfs}
where~$\RR_n$ is convergent and equivalent to~$\EE_0$.
\end{example}

\noindent
We have formalized 
soundness of recording completion
in \isafor \cite{TS09b} (see \texttt{Completion.thy}).
\begin{theorem}
\label{completion thm}
Every successful run of recording completion is sound.
\qed
\end{theorem}

\paragraph{Compare.}
Let $(\EE,\varnothing,\HH_0) \cstep^n (\varnothing,\RR,\HH_n)$ be a
successful run of recording completion and $s \approx t$ an equation.
In the \phase{compare} phase we test joinability of the terms $s$ and $t$
with respect to $\RR$.
If the two terms are joinable, then $s \approx t$ follows
from~\EE and the next phase constructs an \EE-conversion $s \convert^*_\EE
t$. Otherwise, $s \not\approx t$ w.r.t.~\EE.
The \emph{compare} phase is sound 
(cf.\ \rTH{completion thm}).

\paragraph{Recall.}
Let $(\EE,\varnothing,\HH_0) \cstep^n (\varnothing,\RR,\HH_n)$ be a
run of recording completion. Then the \phase{recall} phase transforms a join
$s \to^*_\RR \cdot \FromTB{*}{\RR} t$ into a conversion $s \convert^*_\EE t$ as
follows. For each step $t_1 \ito{i} t_2$ where the index $i$ is not in $\EE$
the corresponding history entry is \emph{inserted}. 
Let $i : \ell \to r$ be the rule with index $i$. Then there must be
a history entry 
$\smash{i : \ell \stackrel{j}{\circ_1} u \stackrel{k}{\circ_2} r}$,
a position $p$, and a substitution $\sigma$ such that
$t_1|_p = \ell\sigma$ and $t_2|_p = r\sigma$. The step 
$t_1 \ito{i} t_2$ is replaced by the conversion
$\smash{t_1 \stackrel{j}{\circ_1} t_1[u\sigma]_p \stackrel{k}{\circ_2} t_2}$.
This process terminates since $i > j,k$, i.e., any history entry (not
in~$\HH_0$) refers
to smaller indices and finally we arrive at a conversion using
indices from~$\EE$.

The next lemma states the desired property of the \phase{recall} phase. Note that we
do not need a \emph{successful} run of recording completion but any 
join $s \to^*_\RR \cdot \FromTB{*}{\RR} t$ is transformed
into $s \convert^*_\EE t$.

\begin{lemma}
\label{LEM:sound}
Let $(\EE,\varnothing,\HH_0) \cstep^n (\EE_n,\RR,\HH_n)$
be a run of recording completion.
Then the recall phase transforms any join
using rules from $\RR$ into a conversion using rules from $\EE$. 
\qed
\end{lemma}

\noindent
Alternatively, one can ensure
${\convert^*_\RR} \subseteq {\convert^*_\EE}$ to derive $s \convert^*_\EE t$ from $s \to^*_\RR \cdot \FromTB{*}{\RR} t$. The former can be established by showing
that all history entries $i\colon s'\smash{\stackrel{j}{\circ_1}}u\smash{\stackrel{k}{\circ_2}}t'$
are consequences of $\EE$ (i.e., $s' \convert^*_\EE t'$) and can thus be used as 
\auxeqs. To avoid cyclic references, history
entries are processed in order of their indices. 
This approach requires the certifier to support
such \auxeqs. In return,
proofs become
much shorter as the history itself is the proof of
${\convert^*_\RR} \subseteq {\convert^*_\EE}$
which obviously has linear size.
In contrast, the recall phase might produce certificates where the conversion
$s \convert^*_\EE t$ is exponentially larger than the join
$s \to^*_\RR \cdot \FromTB{*}{\RR} t$.

\section{Formalization and Certification}
\label{SEC:fac}

To maximize the reliability of the computed results, we have
developed a verified certifier using the proof assistant
Isabelle/HOL.
Based on \isafor~\cite{TS09b} the code generation facilities of
Isabelle/HOL 
allow to generate the verified
certifier \ceta, which is able to certify or falsify conversions, 
completion proofs, and equational proofs and disproofs which are
performed via completion.%
\footnote{Both \isafor and \ceta are freely available from
\url{http://cl-informatik.uibk.ac.at/software/ceta/}.}
For the latter, although \rTH{completion thm} has been formalized,
it is not checked whether the completion rules are applied correctly. 
Instead it is just verified if the result of the completion
procedure is a 
convergent TRS 
equivalent to the initial set of equations. 

To decide
whether $s \convert^*_\EE t$ holds 
it suffices
to find a convergent TRS $\RR$ that is equivalent to $\EE$ and decide whether the
$\RR$-normal forms
of $s$ and $t$ coincide.

For equivalence of $\RR$ and $\EE$ we have to consider two directions.
To decide ${\convert^*_\EE} \subseteq {\convert^*_\RR}$, by convergence
of $\RR$ we just have to
check that for all $s \approx t \in \EE$, the $\RR$-normal forms of $s$ and $t$
coincide.
For the other direction, ${\convert^*_\RR} \subseteq {\convert^*_\EE}$,
we have to guarantee $\ell \convert^*_\EE r$ for all $\ell \to r \in \RR$.
Here, we use the information from recording completion to
get the required derivations. 

Hence, to certify that such a proof is correct we have to guarantee that
$\RR$ is convergent by showing
termination and local confluence.
Concerning termination, already several techniques have been formalized in
\isafor.
Hence, the new part is on the certification of local confluence.
Here, the
key technique is the critical pair theorem
of Huet~\cite{H80}---making a result by Knuth and Bendix~\cite{KB70}
explicit. 
It states that $\RR$ is locally confluent \IFF all critical pairs of $\RR$ are joinable.

During the formalization we detected that in general (no assumption on
the set of variables~$\VV$) there is a problem of renaming variables in
rules for building critical pairs.
To solve this problem without demanding an infinite set of variables, 
we see two alternatives: Either keep the set of variables and 
when building critical pairs 
try to rename variables apart as good as possible; or use an enlarged set of
variables in the definition of 
critical pairs (so that there are enough
variables to perform renamings). 
It turns out that for both alternatives the critical pair theorem does not
hold. 

For the first alternative it is easy to see that joinability of critical
pairs does not imply local confluence. To this end, consider $\VV = \{x\}$
and $\RR = \{\fs f(\fs a,x) \to \fs a, 
\fs f(x,\fs b) \to \fs b\}$.  This TRS is not locally confluent due to the peak
$\fs a \gets \fs f(\fs a,\fs b) \to \fs b$. But without changing $\VV$ it is not
possible to rename the variables of the two rules in $\RR$ apart, such that
their left-hand sides are unifiable. Hence, for the first alternative
all critical pairs are joinable.

For the second alternative, $\RR$ may be locally
confluent although not every critical pair is joinable: 
the next example shows that if $\VV$ is finite and $\RR$ is locally confluent,
then it need not be the case that all critical pairs of $\RR$ are joinable.

\begin{example}\label{confluence example}
Let $\RR_c = \bigcup_{i=1}^4 \RR_i$ be the TRS over $\FF = \{\fs{f},\fs{g},\fs{h},\fs{c}\}$ and $\VV
= \{x_1,\ldots,x_5\}$ depicted in \rTB{TAB:RRc}.
\begin{table}
\begin{center}
$\begin{array}[t]{r@{~}l}
\multicolumn{2}{c}{\RR_1}\\
  \fs{f}(\fs{g}(x_1,x_2),\fs{g}(x_3,x_4)) & \to \fs{h}(x_1,\fs{h}(x_2,\fs{g}(x_3,x_4))) \\
  \fs{f}(\fs{g}(\fs{g}(x_1,x_2),\fs{g}(x_3,x_4)),x_5) & \to \fs{h}(x_5,\fs{h}(\fs{g}(x_1,x_2),\fs{g}(x_3,x_4)))
\end{array}$
\\[1ex]
$\begin{array}[t]{r@{~}l}
\multicolumn{2}{c}{\RR_2}\\
  \fs{h}(\fs{g}(t,x_1),\fs{h}(x_2,x_3)) & \to \fs{c} \\
  \fs{h}(\fs{g}(x_1,t),\fs{h}(x_2,x_3)) & \to \fs{c} \\
  \fs{h}(x_1,\fs{h}(\fs{g}(t,x_2),x_3)) & \to \fs{c} \\
  \fs{h}(x_1,\fs{h}(\fs{g}(x_2,t),x_3)) & \to \fs{c} \\
  \fs{h}(x_1,\fs{h}(x_2,\fs{g}(t,x_3))) & \to \fs{c} \\
  \fs{h}(x_1,\fs{h}(x_2,\fs{g}(x_3,t))) & \to \fs{c} 
\end{array}$
\quad
$\begin{array}[t]{r@{~}l}
\multicolumn{2}{c}{\RR_3}\\
  \fs{h}(\fs{g}(y,x_1),\fs{h}(\fs{g}(x_2,x_3),\fs{g}(x_4,x_5))) & \to \fs{c} \\
  \fs{h}(\fs{g}(x_1,y),\fs{h}(\fs{g}(x_2,x_3),\fs{g}(x_4,x_5))) & \to \fs{c} \\
  \fs{h}(\fs{g}(x_1,x_2),\fs{h}(\fs{g}(y,x_3),\fs{g}(x_4,x_5))) & \to \fs{c} \\
  \fs{h}(\fs{g}(x_1,x_2),\fs{h}(\fs{g}(x_3,y),\fs{g}(x_4,x_5))) & \to \fs{c} \\
  \fs{h}(\fs{g}(x_1,x_2),\fs{h}(\fs{g}(x_3,x_4),\fs{g}(y,x_5))) & \to \fs{c} \\
  \fs{h}(\fs{g}(x_1,x_2),\fs{h}(\fs{g}(x_3,x_4),\fs{g}(x_5,y))) & \to \fs{c}
\end{array}$
\quad
$\begin{array}[t]{r@{~}l}
\multicolumn{2}{c}{\RR_4}\\
  \fs{h}(x_1,\fs{c}) &\to \fs{c}
\end{array}$
\end{center}
\caption{\label{TAB:RRc}Rule schema for $\RR_c$ with $y \in \VV$
  and $t \in \{\fs{c}, \fs{f}(x_4, x_5), \fs{g}(x_4, x_5), \fs{h}(x_4, x_5)\}$.}%
\vspace{-2ex}
\end{table}
It is constructed in such a way that each term
$\fs{h}(\fs{g}(t_1,t_2),\fs{h}(\fs{g}(t_3,t_4),\fs{g}(t_5,t_6)))$ can be reduced to
$\fs{c}$
(via $\RR_2$ if some $t_i$ is not a variable and via $\RR_3$ if $t_i = t_j$ for
$i < j$).
Since there are only five different variables in $\VV$,
indeed every term $\fs{h}(\fs{g}(t_1,t_2),\fs{h}(\fs{g}(t_3,t_4),\fs{g}(t_5,t_6)))$ can
be reduced to $\fs{c}$.
Moreover, all critical pairs, for which one of the rules is taken from 
$\RR_c \setminus \RR_1$, are joinable. 
Hence, the only critical pair that remains to be considered arises between the two rules of $\RR_1$
where $u = \fs{f}(\fs{g}(\fs{g}(x_1,x_2),\fs{g}(x_3,x_4)),\fs{g}(x_5,x_6))$:
\[
\fs{h}(\fs{g}(x_1,x_2),\fs{h}(\fs{g}(x_3,x_4),\fs{g}(x_5,x_6))) \gets  u \to \fs{h}(\fs{g}(x_5,x_6),\fs{h}(\fs{g}(x_1,x_2),\fs{g}(x_3,x_4)))
\]
This critical pair is not joinable, as both terms are $\RR_c$-normal forms.
However, $\RR_c$ is confluent since every \emph{instance} of the critical pair
(w.r.t.\ $\TT(\FF,\VV)$) is joinable
to 
$\fs{c}$.
\end{example} 

The example shows that confluence depends on the set of \emph{variables}
which most often is assumed
to be infinite.
Without this assumption, the requirement that all critical pairs have to be joinable
can be too strict.\footnote{We have only shown this result for
$|\VV| = 5$. However, $\RR_c$ can be adapted to any finite~$\VV$ with $5 \leq |\VV|$.}
Another important consequence is that in the case of finite~$\VV$, Toyama's
modularity result for confluence \cite{Toyama87} does no longer hold.

\begin{corollary}
Confluence is not a modular property of TRSs for an arbitrary set of variables.
\end{corollary}

To summarize, it is not possible to formalize the critical pair theorem for
arbitrary sets~$\VV$.
Hence, we formalized it for strings, where it is conveniently possible
to rename variables of rules apart without changing the type of variables
(by using different prefixes). Of course, if $\VV$ is infinite
we can always obtain a renaming function (take \emph{some} fresh variables) by
the \emph{Axiom of Choice}. However, then
the definition of critical pairs is not executable.

\begin{theorem}
A TRS over $\TT(\FF,\mathit{String})$ is locally confluent \IFF all
critical pairs are joinable.
\end{theorem}

Note that the theorem does not require any variable-condition for
$\RR$. Hence, $\RR$ may, e.g., contain left-hand sides which
are variables or free variables in right-hand sides.

\section{Implementation and Conclusion}
\label{SEC:imp}

We performed experiments%
\footnote{\url{http://cl-informatik.uibk.ac.at/software/kbcv/experiments/12iwc}}
for completion proofs using \KBCV~\cite{SZ12} and \MKBTT~\cite{MKBTT_new}
(on 115 ESs).%
\footnote{\url{http://cl-informatik.uibk.ac.at/software/mkbtt}}
Within a time limit of 300 seconds,
\KBCV could complete 86 ESs and \MKBTT 80 ESs while both tools together
succeeded on 94.
The corresponding 94 completion proofs could be certified by \ceta (version 2.4).
For an evaluation of other completion tools we refer to~\cite{KH11}.

In our experiments we considered both possibilities (mentioned at the end
of Section~\ref{SEC:rec}) to ensure
${\convert^*_\RR} \subseteq {\convert^*_\EE}$.
While $\KBCV$~1.6 performs the recall phase to explicitly
construct $\ell \convert^*_\EE r$ for each $\ell \to r \in \RR$, 
$\KBCV$~1.7 just exports the relevant history entries, which are used as
\auxeqs.
Hence it is not surprising that from the 86 ESs which \KBCV~1.6 could
complete only 80 have been certified. For two ESs 
(\file{TPTP_GRP487-1_theory} and 
 \file{TPTP_GRP_490-1_theory})
the recall phase did not terminate within the time limit
and for the remaining ESs 
(\file{LS94_P1},
 \file{TPTP_GRP_481-1_theory},
 \file{TPTP_GRP_486-1_theory},
 \file{TPTP_GRP_490-1_theory})
the certificate was too large 
(365\,MB, 230\,MB, 406\,MB, 581\,MB)
for \ceta.
However, when using \auxeqs all proofs could be computed and
certified (typically within a
second).
Hence 
further optimization of the proof
format seems dispensable.

While~\MKBTT follows recording completion, \CIME implements an annotated
version of ordered completion~\cite{CC05}. Here---in contrast to our
approach---the history is not saved as a stand-alone component but directly
integrated into terms, equations, and rules. Hence a term $t$ comes with an
original version $t^0$, a current version $t^*$, and a reduction sequence from
$t^0$ to $t^*$. Similarly an equation $s \approx t$ also contains all
intermediate (rewrite) steps that show that both terms are equal. It
requires further investigations to evaluate the pros and cons of the two
approaches.

\paragraph{Acknowledgments:}
We thank Sarah Winkler for integrating certifiable output into \MKBTT and
helpful discussion.

\bibliographystyle{plain}
\bibliography{iwc}

\end{document}